\documentstyle[12pt,epsfig]{article}

\def\aj{{\it AJ}~}			
\def\araa{{\it ARA\&A}~}		
\def\apj{{\it ApJ}~}

\def\mnras{{\it MNRAS}~}

\def\prd{{\it Phys.~Rev.~D}~}	
	
\def\prl{{\it Phys.~Rev.~Lett.}~}

\def\pnas{{\it Proc.~Nat.~Acad.~Sci.}~}

\def\mxth{\mathsurround=0pt }
\def\xversim#1#2{\lower2.pt\vbox{\baselineskip0pt \lineskip-.2pt
    \ialign{$\mxth#1\hfil##\hfil$\crcr#2\crcr\sim\crcr}}}
    \def\lesssim{\mathrel{\mathpalette\xversim <}}

\input epsf
\begin{document}
\title{The Cosmic Triangle: Assessing the State of the Universe }

\author{Neta A. Bahcall$^1$, Jeremiah P. Ostriker$^1$\\
 Saul Perlmutter$^2$ and Paul J. Steinhardt$^3$}

\maketitle


\vspace{.1in}
\noindent
The cosmic triangle is introduced as a way of representing
the past, present, and future 
status of the universe.
 Our current location within the cosmic  triangle
is determined by the answers to three questions:
How  much matter is in the universe? Is the expansion rate
slowing down or speeding up? And, is the universe flat?
A  review of recent observations  suggests 
a universe that is lightweight
(matter density about one-third the critical value),
is accelerating, and is flat. The acceleration implies the
existence  of  cosmic dark energy that overcomes
the gravitational self-attraction of matter and causes the
expansion to speed up.
\vspace{.2in}

\noindent
$^1$Princeton University Observatory, Princeton, NJ 08544  \\
$^2$Institute for Nuclear and Particle Astrophysics, E.O. Lawrence
Berkeley National Laboratory, Berkeley, CA  94720
\\
$^3$Department of Physics, Princeton University, Princeton, NJ 08544

 \newpage
As the new millennium approaches, 
novel technologies are opening new windows on the universe.
Whereas previously we relied primarily  on fossil evidence 
found in the local neighborhood of our galaxy  to infer
the
history of the universe, now we  can
 see
directly the evolution of the universe
over the past  15 billion years extending  as far back as   
  100,000 years after the big bang.
 Thus far,  the  picture of the past history of 
 the cosmos has altered
only slightly: the new observations described in this 
paper are perfectly consistent with the standard big bang 
picture of
the expansion of the universe from a hot, dense gas, 
the synthesis of the
elements in the
first few minutes, and the  growth of 
structure through gravitational amplification of small,
initial inhomogeneities.
 However, the  expectation 
for the future  has
been
dramatically  revised.  Based on the conventional
assumption that the universe
contains
only  matter and radiation -- the forms of energy we can
readily detect -- 
the expectation for  the future
had been  that the expansion rate of the universe 
would slow continuously
due to the
gravitational self-attraction of matter.  
The only major issue
seemed to be 
whether
the universe would expand forever or ultimately recollapse to a big
crunch.
Now, the mounting evidence described below
is forcing us to consider the possibility
that some cosmic dark energy  exists that opposes the 
self-attraction of
matter and is causing the expansion of the universe to accelerate.
\\

Since the discovery of cosmic expansion by Hubble and
Slipher~\cite{Hubble20} in the 1920's, the standard
assumption had been that all energy in the universe is in the form of
radiation and ordinary matter (electrons, protons, neutrons,
and neutrinos -- with mass counting as energy at the rate $E = mc^2$).
Over the next several decades, though, theory concerning the stability
of galaxies~\cite{OP73}, 
observations of the motion of galaxies in 
clusters~\cite{Dark1,Bahcall77} and  stars and gas surrounding
galaxies~\cite{B39,Dark2} indicated that most of the
mass in the  universe is dark 
and does not emit or absorb light~\cite{ML1,Trimble87}.
In the 1980's, the proposal of dark matter found resonance in the
``inflationary universe''
scenario~\cite{Gut,Lin1,Lin2,Davies}, a theory 
of the first $10^{-30}$~seconds designed to address several
questions left unanswered by
the big bang model:  Why is the universe so homogeneous and 
isotropic?  Why is the curvature of space so insignificant?
Where did the initial inhomogeneities come from that give rise
to the formation of  structure~\cite{G4,G41,G42,G43}? 
The standard 
inflationary theory
predicts that the universe is spatially flat;
according to Einstein's theory of general relativity, 
this  fixes
 the total energy density
of the universe  to equal precisely the critical value, 
$\rho_c \equiv 3H_0^2/8 \pi G \sim 1.7 \times
10^{-29}$~g~cm$^{-3}$, where $H_0$ is the current value of
the Hubble parameter and $G$ is Newton's gravitational constant
(see 
Eq. (1)).
Measurements
show that ordinary matter
 and radiation account for less than 10\%
of the predicted
value~\cite{Walker91,BurlesTytler97a,SchrammTurner98}. 
Inflation thus  seemed to call for dark matter.
The observational case for dark matter continued to grow, 
as discussed below,
and
particle physicists proposed various hypothetical particles,
motivated by supersymmetry and unified theories, that could reasonably
explain the dark matter.  The new consensus model became
the ``cold dark matter'' picture which predicts that the universe
contains  primarily
cold, nonbaryonic dark matter~\cite{Cold,White87,footcold}.  
Although the total mass density
identified by observations
still fell short of the critical
value~\cite{ML1,Trimble87,Gott74,Bahcall92}, 
many cosmologists adopted as a working hypothesis 
the critical density
model, trusting that
something would fill the gap.
\\

The last few years have seen signs of
another shake-up of the standard model~\cite{Kofman93,OS95,KT95}.
First, improved observations confirmed that the total mass
density is  probably less than half of the
critical density~\cite{Bahcall95,Carlberg96,Bahcall98}.
At the same time, combined 
measurements of the cosmic microwave background
(CMB)
temperature fluctuations and the distribution of galaxies on
large scales began to suggest
that the universe may  be flat~\cite{OS95}, consistent with
the standard inflationary prediction.  The only way to have a
low mass density and a flat universe, as expected from 
the inflationary theory,  is if
an additional,  nonluminous,  ``dark'' energy component dominates
the universe today.  
The dark energy would have to resist  gravitational collapse,
or else it would already have been detected as part of the clustered
energy in the halos of  galaxies.
But, as long as most of the energy of the
universe resists gravitational collapse, it is impossible for
 structure in the universe to form.
The  dilemma can  be resolved if the hypothetical
dark energy was negligible in the past and then over time
became the dominant energy in the universe. 
According to general relativity \cite{footnote1}, this requires
that the dark energy have a remarkable feature: {\it negative}
pressure.
This  argument~\cite{OS95} would rule out almost all of
the usual suspects, such as cold dark matter, neutrinos,
radiation,  and kinetic energy, because they have zero or positive
pressure.

With the recent measurements of distant exploding stars, supernovae,
the existence of  negative-pressure dark energy has
begun to gain broad consideration.  Using 
Type Ia supernovae as standard candles
to gauge the expansion of the universe,
observers  have found evidence that the universe is
 accelerating~\cite{Perlmutter98,Riess98}.
A dark energy with significant
negative pressure~\cite{OS95,Peeb84} will in fact
cause the expansion of the universe to speed up, so the supernova
observations provide  empirical evidence of a dark energy with
 strongly negative 
pressure~\cite{Perlmutter98,Garnavich98,Wang99,Perlmutter99}.

The news has brought the return of
the cosmological constant,
first introduced by Einstein 
 for the purpose of allowing a static universe with the
repulsive cosmological constant delicately
balancing the gravitational
attraction of matter~\cite{Einstein17}.  In its present incarnation, the
cosmological constant is out of balance, causing the
expansion of the universe to accelerate.  It can be viewed as
a vacuum energy assigned to empty space
itself,   a form of  energy with negative
pressure.
Cosmologists are familiar with other hypothetical forms of
dark energy with negative pressure that
can accelerate the universe.
In inflationary cosmology, a cosmic field, similar
to an electric field in the sense that  it  pervades space 
and assigns a field value and energy to each point in it;
acceleration is caused by a  cosmic
field whose kinetic energy 
is much less than its potential energy~\cite{Lin1,Lin2}.
A different field, with much tinier energy, coined
``quintessence"~\cite{CDS98}, 
could account for the acceleration
suggested to be observed today. 
Unlike a cosmological constant, quintessence energy
changes  with time and naturally 
develops inhomogeneities that can produce variations in the 
distribution of mass  and  the CMB
  temperature observed today~\cite{CDS98,Ratra88}.

The introduction of a dark energy component which leads to the
acceleration
of the universe is neither a simple nor modest change.
New challenges to cosmology and to fundamental physics are immediately
posed.  And the future of the
universe  is totally altered.
Although it is premature to call the new evidence for an accelerating
expansion and exotic dark energy conclusive,
the case is strong enough and the implications dramatic enough
that the time is ripe for a  new report on the state of the universe.

\begin{center}{\bf The Cosmic Triangle}
\end{center}

According to Einstein's theory of general relativity, the evolution of
the universe  is determined by
the forms of energy it contains and the curvature of space.
Einstein's equations can be reduced to a simple form known
as the  Friedmann equation:
\begin{equation} \label{ein}
H^2 = \frac{8 \pi G}{3} \rho - \frac{k}{a^2}.
\end{equation}
 On the left-hand-side is the Hubble parameter
 $H = H(t)$, which measures the expansion rate of the universe
as a function of time.
The current value, $H_0$, is  $65 \pm 10$~km~sec$^{-1}$~Mpc$^{-1}$,
where one megaparsec (Mpc) is $3.26 \times 10^6$~light
 years~\cite{Freedman98,TammannConf98,MyersSZE97}.
 The right-hand side contains the factors which determine the
 expansion rate.  The first factor  is the energy density $\rho$ (times
Newton's gravitational constant $G$). The energy density $\rho =
\rho(t)$
 can have several
different subcomponents:
a mass density associated with ordinary and dark matter,
the kinetic energy of the particles and radiation, the energy
associated with fields (such as quintessence), and the
vacuum energy density or, equivalently,  the cosmological constant.

The second term on the right-hand side describes the effect of
curvature of space on the expansion of the universe. The curvature
constant $k$ can
be positive, negative or zero.  The parameter $a = a(t)$, known as the
scale factor, measures how much the universe stretches as a function of
time.
It can be thought of as proportional to the average distance between
galaxies.
As the universe stretches, the curvature is diminished, as indicated in
the equation.
The terms closed, open and flat  refer, by definition,  to the cases of
positive, negative, and zero curvature.  It has been common to use the
same terms to
describe whether the universe will ultimately recollapse, expand
forever, or lie on the
border between expansion and recollapse.  This second usage
does not
necessarily apply if there is vacuum density or quintessence, a point
which often causes
confusion.  For example, if there is vacuum energy,
it is possible to have a universe that is
closed (positive curvature), but
which expands forever because the acceleration 
due to the cosmological
constant
overcomes the curvature effect \cite{over} which 
would otherwise bring the expansion
to  a halt and then recollapse.  

For simplicity, we will consider a universe composed today 
of baryonic
(ordinary) and
dark (exotic) matter, curvature and vacuum energy ({\it i.e.},
a cosmological constant, $\Lambda$).  The fractional
contributions
to the right-hand side of the Friedmann equation, which depend on the
relative values of the matter density, vacuum energy density ($\rho_\Lambda$)
and curvature, are given the
symbols $\Omega_m \equiv 8\pi G \rho_{\rm matter} /(3H^2)$,
$\Omega_{\Lambda} \equiv 8\pi G \rho_\Lambda /(3H^2) \equiv \Lambda/(3
H^2)$, and $\Omega_k \equiv -k/(aH)^2$, respectively~\cite{foot30}.
Dividing both sides of Eq.~(\ref{ein}) by $H^2$ yields a simple sum rule

\begin{equation} \label{sum}
1= \Omega_m + \Omega_k + \Omega_{\Lambda}.
\end{equation}
For any other energy component, such as 
quintessence, a  term $\Omega_Q$
would be
added to the right-hand side of Eq.~(2).
The sum rule can be represented by an equilateral
triangle (Fig. 1).  Lines of
constant $\Omega_m$, $\Omega_k$ and $\Omega_{\Lambda}$ run
parallel to each of the edges of the equilateral triangle.
Every point lies at an intersection of lines of constant
$\Omega_m$, $\Omega_k$ and $\Omega_{\Lambda}$ such that the sum
rule is satisfied.
Although $\Omega_m$ is non-negative, the curvature and cosmological
constant can be positive or negative. 

Inflationary theory~\cite{Gut,Lin1,Lin2,Davies} proposes that
the universe underwent a brief epoch 
of extraordinary expansion during the first $10^{-30}$ seconds
after the big
bang which ironed out the curvature, setting $\Omega_k = 0$.
If the curvature is zero, i.e., the universe is flat,
then the sum rule reduces to $\Omega_m+  \Omega_{\Lambda} =1$,
corresponding to the blue line (marked ``flat") in the figure.
The yellow line indicates the division between models in 
which the  expansion rate  is currently decelerating versus accelerating.
The competition between the decelerating effect
of the mass density and the accelerating effect of
the vacuum energy
density 
can
be understood from Einstein's equation for the
stretching of the scale factor $a(t)$:
\begin{equation} \label{Ein}
\ddot{a} = -\frac{4 \pi G}{3} (\rho+ 3 p) a,
\end{equation}
where $p$ is the pressure associated with whatever energy is
contained within the
universe.    If the universe
contains ordinary matter and radiation, then  $\rho+ 3p$
is positive, and the expansion decelerates,
$\ddot{a} <0$.
However, exotic components like vacuum-energy  and
quintessence~\cite{CDS98,Ratra88} have 
sufficient {\it negative} pressure  to make $\rho+ 3p$ negative,
inducing cosmic acceleration.

Finally,
models of special interest have been highlighted in the Figure;
they form nearly an equilateral triangle of their own.
The standard cold dark matter model
({\sc Scdm}), the most simple possibility,  has
$\Omega_m =1$ and no curvature or vacuum component. The
model assumes a ``scale-invariant" 
spectrum of initial density fluctuations, a spectrum in which 
the magnitude of the inhomogeneity
is the same on all length scales, as predicted by standard
inflationary cosmology~\cite{G4,G41,G42,G43}.
A model that  better fits the observations and  which retains the 
simple condition of $\Omega_m=1$ is the ``tilted" {\sc Tcdm}
in which the fluctuation spectrum is tilted so that the
average inhomogeneity increases with length scale,
unlike the standard  inflationary 
prediction. The open cold dark matter 
({\sc Ocdm})  model has low mass density and 
and no vacuum component; the best-fit  version has
a mixture of one-third matter density and
two-thirds curvature (but no vacuum energy) 
with a   spectrum that is tilted the opposite way from 
the {\sc Tcdm} model (the inhomogeneity decreases
as the length scale increases).  As an example of a dark 
energy plus cold dark matter model ($\Lambda${\sc cdm}), 
a current best estimate model, we will consider 
a mixture of two-thirds vacuum density
(or cosmological constant, $\Lambda$)
 and one-third matter density
(but no curvature) and the  standard untilted spectrum predicted
by inflationary theory.
 The parameters for each of the four models, shown in Table 
 I, has been  chosen to 
 best fit for each type of model to
 the current observational constraints discussed below.
All the models (and our analysis) assume the standard inflationary
prediction that the density fluctuations are gaussian 
and adiabatic ({\it i.e.}, radiation, 
ordinary and dark  matter fluctuate spatially in the same 
manner)~\cite{G4,G41,G42,G43}, which agree  with current 
observations.
The age of the universe 
(in units of 10$^9$~y)
is consistent with the most recent estimates of
the ages  of the oldest
  stars~\cite{Chaboyer98,Salaris98}.

The models may be distinguished observationally   
by answering three fundamental questions:
Is there enough matter to close (flatten) the universe?
Is the expansion rate accelerating, providing evidence for a new
dark energy?  Is the universe curved?
In the next sections we describe a series  of independent
tests aimed at addressing these three questions.  
The best constraint for each question is  represented
as a strip in the plot in Fig.~2.
Together, these constraints determine our location
in the cosmic triangle plot and, thereby, the past and 
future evolution of the universe.

\begin{center}
{\bf Is There Enough Mass to Close the Universe?}
\end{center}

The consensus for a low-mass-density universe ($\Omega_m <1$)
has been building slowly for over a decade, although, truth be said,
there was never credible 
evidence otherwise~\cite{ML1,Trimble87,Gott74,Bahcall92,OS95,Bahcall95,Carlberg96,Bahcall98}.  
The determination of the
universe's mass density is  currently the best-studied
of the three cosmological parameters, and is supported by
a number of independent measurements.
 Although each observation has its
 strengths, weaknesses  and assumptions,
 they all indicate that $\Omega_m<1$.

\noindent
{\it Mass-to-light method.}
One of the oldest and simplest techniques for estimating the total mass
of
the universe entails a two-step process:
first determine the average ratio of  the
mass to the emitted light of the largest systems possible;
then,  multiply by the total measured luminosity
density of the universe.  This totals up all the
mass associated with light to the largest scales.
Rich clusters of galaxies are the  largest (1-2 Mpc in radius)
and most
massive ($2-10 \times 10^{14}$ solar masses within 1 Mpc)
bound systems known
for which mass has been reliably measured.
Nowadays,
cluster mass can be inferred from three independent methods:
the galaxy motion within the cluster,
 the temperature of the hot intracluster gas, and  gravitational
lensing by the cluster mass (the  distortion
of background galaxies' images by the cluster's
gravitational potential).
There is good agreement among these independent estimators.
The mean cluster mass-to-light ratio ($M/L$),
about  $200 \pm 70$ times the
mass-to-light ratio for the sun, indicates that there is a great
deal of dark matter within clusters~\cite{Bahcall95,Carlberg96}.
Nevertheless, even if we assume that all light in the universe
is emitted from
objects that have as much mass per unit light as clusters,
the total mass would not be sufficient to close
the universe.
Multiplying   $M/L$  by the observed
luminosity density, one obtains
$\Omega_m =0.2 \pm 0.1$~\cite{ML1,Bahcall95,Carlberg96}.
Recent studies of the dependence of $M/L$ on
scale indicate 
that $M/L$ is nearly  constant on  large
scales ranging up to
supercluster size (10 Mpc),
suggesting no additional dark matter is tucked away on large
scales~\cite{Bahcall95,Kaiser98}. 
\\

\noindent
{\it Baryon fraction method.}
An independent method of estimating the mass of the universe,
 also based on rich clusters, entails measuring the ratio of the
baryonic  to  total mass in clusters~\cite{WNES93,White95}.  
Because
 clusters  form through
gravitational collapse, they scoop up
the mass over a  large  volume of space such that
the ratio of baryons to total matter in the collapsed cluster
should be representative of the cosmic average to within
20\%~\cite{Lubin96,Arnaud98}.
The big bang model of primordial nucleosynthesis constrains
 the baryon density to be  $\Omega_b = 0.045 \pm 0.0025 $
(based on the
cosmic abundance of helium and deuterium, and using 
$H_0=65$~km~$s^{-1}$~Mpc$^{-1}$)~\cite{Walker91,BurlesTytler97a,SchrammTurner98}. 
Thus, if one can measure
 the average  baryon ratio in the universe,
  $\Omega_b/\Omega_m$, it can be used
  with the known $\Omega_b$
 to determine $\Omega_m$.
 A cluster's baryon ratio can be determined from the baryonic mass in
the cluster (obtained by measuring
the x-ray emission from the hot intracluster gas and adding the
mass of the stars~\cite{WNES93,FHP97}) divided by the total cluster mass.
The baryon ratio is found to be $\Omega_b/\Omega_m \approx 0.15$,
much larger than the 0.045 value expected if
$\Omega_m=1$~\cite{WNES93,White95,Lubin96,Arnaud98}.
The observed ratio
 corresponds to a mass density 
 of $\Omega_m=0.3 \pm 0.1$.
If some  baryons are ejected from the cluster during
gravitational collapse, as suggested by cosmological simulations
\cite{Lubin96,Arnaud98},
or if some baryons are bound in nonluminous objects such as 
rocks  or planetary-sized objects,
then the actual value of $\Omega_m$ is lower than this estimate.
\\

\noindent
{\it Cluster Abundance and Its Evolution.}
A third feature of rich clusters which constrains $\Omega_m$
is the number density of clusters as a function of cosmic time (or red
shift)~\cite{Bahcall98,Eke96,Bahcall97,Carlberg97,H97,Wang98}.
Rich clusters are the most recently formed gravitationally bound
objects in the universe.  
The observed present-day $(z \sim 0)$ cluster abundance
provides a strong constraint on the normalization of the power spectrum
of density fluctuations-- the seeds that created the clusters--
on the relevant cluster scales (see next
subsection)~\cite{Bahcall92,Eke96,White93}. The 
$\Lambda${\sc cdm} and {\sc Ocdm} models 
are consistent with the observed cluster
abundance at $z \sim 0$.
{\sc Scdm}, however, when normalized to
match the observed fluctuations in the CMB
(see next subsection), produces too many clusters
at all red shifts (Fig. 3)~\cite{Bahcall92,Bahcall97,White93,Ost}.  
  The
  {\sc Tcdm} model preserves $\Omega_m=1$ and more nearly
  fits the present day cluster abundance (Fig. 3, 4).

 The evolution of cluster
abundance with red shift breaks the $z=0$  degeneracy among the
models~\cite{Bahcall98,Eke96,Bahcall97,Carlberg97}.  
The $\Omega_m <1$ models  ($\Lambda${\sc cdm} and
and {\sc Ocdm})  predict relatively little change in
the number density of rich clusters as a function of red shift
because, due to the low matter density,
hardly any structure growth has occurred since $z \sim 1$.
For the $\Omega_m=1$ {\sc Tcdm} model, structure has been growing
 steadily and  rich clusters could only have
formed recently; the number density  of rich clusters
at $z \simeq 0.5 - 1$ is predicted to
be exponentially smaller than today.
 The observation of even one
 massive cluster at high red shift ($z > 0.6$)
 suffices to rule out the $\Omega_m=1$ model.  In fact, 
 three clusters
 have been observed already (Fig. 3), suggesting 
 a low-density
 universe, $\Omega_m=0.25^{+0.15}_{-0.10}$ 
 (1$\sigma$)~\cite{Bahcall98}.
 A caveat for this method is that it assumes that
 the initial spectrum of density perturbations is gaussian, as
 predicted by  inflation, which has not yet  been
 carefully confirmed observationally (but see~\cite{COS98}) on the
 cluster scales.
 \\

\noindent
{\it  Mass Power Spectrum.}
  The mass power spectrum (Fig. 4)  measures  the degree of
  inhomogeneity
  in the universe's mass distribution on different distance scales.
  Beginning from a cosmological model, the mass
  power spectrum depends on the initial spectrum of inhomogeneities
  ({\it e.g.}, the stretched-out
  quantum fluctuations predicted
  by inflation), the recent creation of new perturbations,
  and how those inhomogeneities have evolved over time
  (which depends on the cosmological parameters).
Existing measurements of the present day abundance of galaxy 
clusters
constrain  the mass inhomogeneity on 
the smallest scale
for which the  power spectrum can be reliably interpreted
(about 10~Mpc).  Observations of
temperature fluctuations  in the cosmic microwave background across
the sky, as measured by the COBE satellite~\cite{BunnWhite97},
constrain both the  amplitude and shape of the spectrum
on the largest observable scales.

Galaxy surveys
are beginning to probe  intermediate scales, from
10 to 1000~Mpc (Fig.~4)~\cite{Peac97,SDSS}.
The theoretical
model predicts the distribution of all the mass, while 
  observations of galaxies
  reflect the luminous, baryonic matter only.
   If the
   luminous matter
   follows the total mass,
   the mass distribution is said to be ``unbiased."
   Otherwise, the ratio of overdensity in luminous matter to that
    in the
    total mass
    is termed the ``bias." On small scales, the bias may
    vary with distance scale and local environment.
    These complications can be avoided by focusing on measurements
     of the
power spectrum on large scales
(more than 10~Mpc; Fig. 4)
where the inhomogeneities are small and the bias is expected to be
small~\cite{Fry86,TegmarkPeebles98}.
Although current galaxy measurements are inconclusive, especially
given uncertainty in the bias, future surveys, 
such as the Sloan Digital
Sky Survey~\cite{SDSS},
are poised to test the shape of the spectrum on the intermediate scales.

Inflation predicts the {\it shape} of each
spectrum~\cite{G4,G41,G42,G43}, but it does not predict 
its
{\it normalization} (i.e., amplitude).  The normalization is
determined
from observations, mainly the observed cluster abundance (on 10~Mpc
scales) and
the CMB fluctuations (on 1000~Mpc scales).  The $\Omega_m = 1$ {\sc Scdm} model,
normalized to the CMB fluctuations on large scales, is 
inconsistent with
the
cluster abundance (predicting over ten times more clusters than
observed;
Fig. 3).  {\sc Scdm} is thus inconsistent with
observations~\cite{Bahcall92,OS95,Bahcall98,Eke96,H97,White93,Ost}. The
 model can be ``forced" to agree  approximately with 
both the cluster abundance on small scales and the CMB fluctuations on
large scales by
tilting the power spectrum (by about 30\%) from its standard shape.
This tilted variant of the 
{\sc Scdm} model-- {\sc Tcdm}-- is thus nearly
consistent with
both constraints.  
  The power spectra of the $\Lambda${\sc cdm} and {\sc Ocdm} models can be
normalized so
that they agree with both the CMB and cluster observations (with a
 30\% tilt needed for {\sc Ocdm}).  
 Future observations,  on all scales, will greatly improve
the power
spectrum
constraints.  This will allow a measure of $\Omega_m$ from the shape of
the
spectrum; currently this measure suggests a low 
value of $\Omega_m$, but
with large uncertainty.

\vspace{.1in}
\noindent
{\it  Overall Estimate of $\Omega_m$.} 
The independent methods described above using
clusters of galaxies yield a consistent determination of a low
mass-density
universe, nearly independent of $\Omega_\Lambda$ or $\Omega_k$.  
Other methods discussed later in the paper, 
such as statistics of gravitational lensing,
 large 
scale velocities~\cite{footSW}, and  measurements
of the CMB anisotropy, place additional 
constraints 
on $\Omega_m$ in combination with other 
parameters and assumptions; though less constraining,
these too suggest a low-mass
density~\cite{Wang99,Perlmutter99}. 
The net result  is
represented by the ``clusters" band (1$\sigma$) in the cosmic triangle of
Fig.~2. 
It is remarkable that a single value of $\Omega$, $\Omega_m \simeq 1/3$,
is consistent with so many, diverse observations.

\begin{center}
{\bf Is the Universe's Expansion Accelerating?}
\end{center}

Changes in the cosmic expansion rate can be studied
using the  observed brightness-red shift
relation.  A set of standard candles
(objects of
known luminosity) spread throughout the universe
are used to determine  the relation between distance and red shift.
The distance $d_L$ is determined by comparing the known
luminosity $L$ to the flux observed at  Earth, $f_{obs}$, and
invoking the inverse-square law $(f_{obs}=L/4\pi d^2_L)$.
By studying standard candles at different observed fluxes, we study
objects whose light was emitted at different cosmic times.  The red
shift $z$ of the object measures the expansion of the universe since
that time.

For relatively nearby standard candles the distance $d_L$ is a
simple linear function of red shift, as given by the Hubble relation
of the expanding universe: $ H_0 d_L=cz$, where $c$ is the
speed of light.  However, the linear relation 
is only an approximation.
If we study standard candles farther away,
the non-linearities in the $d_L$-$z$ relation
become important because the universe's expansion may be
decelerating or accelerating.  The results are most sensitive to the
difference between
$\Omega_m$ (which decelerates the expansion) and $\Omega_\Lambda$ (which accelerates the expansion)
and are rather insensitive to the curvature $\Omega_k$.

\noindent {\it Supernovae}.
Type Ia  supernovae are the current best candidates for 
 standard candles. They have the advantage that they are
bright and can be seen at cosmic distances. As a class,
Type Ia supernovae are not {\em all} identically luminous,
but examination of nearby supernovae indicates
that they may be converted into reliable distance indicators
by calibrating them according to the time scale of their brightening and
fading~\cite{Phillips93,Perlmutter97}.
Two efforts are underway to collect  data on the red shift, luminosity
and light curves of distant supernovae, the Supernova Cosmology Project
(SCP)~\cite{Perlmutter98,Perlmutter97,Perlmutter95} and
the High-Z Supernova 
Search (HZS)~\cite{Riess98,Garnavich3SNe,Schmidt98}. By now,
studies of over 50
Type Ia supernovae at  $z=0.3$ to 0.9
have been published and calibrated with a comparable
number of  nearby supernovae~\cite{Hamuy9596,Riess99}
at $z\lesssim 0.1$.

 The results of the two studies
show that the distant supernovae are fainter, thus more distant than
expected for a decelerating universe (Fig.~5).
It appears that the expansion rate
is accelerating, indicating the existence of dark energy with negative
pressure, such as $\Omega_\Lambda$.  The best-fit results 
(Fig. 2) can be approximated
by the linear combination $0.8\Omega_m - 0.6\Omega_\Lambda = -0.2\pm 0.1
(1\sigma)$~\cite{Perlmutter98,Goobar95}.  For a
flat universe ($\Omega_m + \Omega_\Lambda=1$), the best-fit values
are approximately $\Omega_m =0.25 \pm 0.1$ and
$\Omega_\Lambda=0.75 \pm 0.1$ ($1 \sigma$)
for the combined results of  SCP team~\cite{Perlmutter98}  and 
the two analyses 
of the HZS team~\cite{Riess98}.
These values are in excellent agreement with the $\Omega_m$ results
discussed above.  In particular, all
flat $\Omega_m = 1$ models, which are
identical
in their $d_L - z$ predictions, are formally ruled out at the $8
\sigma$
level.

The caveats for this test are possible uncertainties in
the cross-comparison of the near and distant supernovae.
Distant supernovae are calibrated using nearby
supernovae assuming that the lightcurve time scale accounts for
any relevant evolution of Type Ia supernovae.
Although known evolutionary and dust obscuration effects have
been taken into account,
there remains the concern that there are additional evolutionary
or dust effects at large red shift that have not been noted
before.
Further investigations are underway using observations
comparing nearby and distant supernoave.  The current 
results suggest that the expansion of the universe is accelerating,
indicating the existence of 
a cosmological constant or dark energy.

\vspace{.1in}
\noindent
{\it Gravitational Lensing Statistics}.
 Gravitational lensing due to accumulations
of matter along the line of sight to distant light sources provide
another
potentially  sensitive measure of our position in the cosmic triangle.
These measures can be used in two ways. The first method uses the
abundance
of multiply imaged sources such as quasars, lensed
by intervening
galaxies~\cite{TOG84,MaozRix93,Kochanek95,Falco98}. The
probability of finding lensed images is 
directly proportional to the number of galaxies (lenses) along the path
and thus to the distance in light-years to the source. This distance
(for fixed $H_0$) increases dramatically for a
large value of the cosmological constant: 
the age of the universe
and the distance to the galaxy
become large in the presence of $\Omega_{\Lambda}$
because the
universe has been expanding for a longer time (compared
with an $\Omega_m=1$ case); therefore,  more lenses are predicted
if $\Omega_{\Lambda}>0$.
Using this method, an upper limit of $\Omega_{\Lambda} < 0.75$ (95\%CL)
has been obtained~\cite{TOG84,MaozRix93,Kochanek95,Falco98}, 
marginally consistent with the
supernovae results.  The caveats of this powerful method
include
its sensitivity to uncertainties in the number density and lensing
cross-section of the lensing galaxies and the
number density of distant faint quasars.
A second method is lensing by  massive clusters of
galaxies~\cite{W97,Bartelmann98}. Such lensing produces
widely separated 
lensed images of quasars and  distorted images of background
galaxies.
The observed statistics of this lensing, when compared with
numerical simulations, rule out the $\Omega_m = 1$
models~\cite{W97,Bartelmann98} and set an upper bound 
of $\Omega_{\Lambda}<0.7$~\cite{Bartelmann98}. 
The limit is sensitive to the resolution
of the numerical simulations, which are currently improving.

\begin{center}
{\bf Is the Universe Curved?}
\end{center}

The curvature of the universe can  be measured from the
highest-red shift cosmological test, the 
cosmic microwave background.
The  CMB power spectrum provides a measure of the inhomogeneity
in matter and energy at  $z \approx 1000$,
corresponding to a few 100,000 years after the big
bang.  
The 
power spectrum is the 
root-mean-square fluctuation
in the CMB temperature  ---the temperature
``anisotropy''--- as a function of 
the angular scale expressed as
an integer multipole moment, $\ell$.  A given $\ell$
corresponds roughly to angle $\pi/\ell$ radians.
Each cosmological model produces a distinguishable CMB temperature
anisotropy fingerprint~\cite{Steinhardt95,Hu}.
On large angular scales (small $\ell$ values), the CMB spectrum probes
 inhomogeneities which span distances so large ($\sim 1000$~Mpc)
 that neither light nor any other interaction has had time to
 traverse or modify them.  These inhomogeneities are a direct
 reflection of the initial spectrum ({\it e.g.}, as
 created by inflation). 
If the models predict an untilted or a tilted spectrum,
then the  CMB anisotropy spectrum has a plateau that is flat
or tilted, respectively.
On small angular scales (less than a degree or $\ell > 200$),
the anisotropy spectrum has peaks and valleys created by
the small-scale inhomogeneities; on these scales,
there has been 
sufficient time 
for light to traverse them and for the matter to respond
gravitationally to the density fluctuation.
The hot gas of baryons and radiation begin a series of acoustic
oscillations in which matter and radiation are drawn by gravity
into regions of high density and then rebound due to the
finite pressure of the gas. 
On scales corresponding to the ``sound horizon"
(the maximum distance pressure waves can travel from the 
beginning of the universe up to the 
time the CMB is emitted),  the mass has had time to 
undergo maximum collapse around the dense regions
so as to produce maximum anisotropy but
has not had time to rebound. 
Hence, a peak in the power spectrum is anticipated  on the angular
scale corresponding the sound horizon, and this should 
be the peak with the largest angular scale (smallest $\ell$).
An interesting
feature is that the physical 
length corresponding to the sound horizon
is relatively insensitive
to the cosmological model. The angular scale which it
subtends on the sky depends only on the overall
curvature of space; the curvature 
 distorts the path of light so that the sound
horizon appears  bigger or smaller  on the sky depending on 
whether the curvature is positive or negative.
If the universe is flat, the sound horizon subtends
about a degree on the sky (resulting in a power spectrum 
peak near $\ell \approx 200$), whereas the
angular size is smaller in a curved open model
(resulting in a peak near
$\ell \approx 200/\sqrt{\Omega_m + \Omega_{\Lambda}}$)~\cite{Kam}.

The CMB anisotropy was first detected by
the COBE (Cosmic Background Explorer) satellite in 1992~\cite{S92},
followed by a series of ground- and
balloon--based experiments~\cite{MSAM,QMAPa,SK,CAT,RING}.
Here we have selected published  experiments 
which
measure at several frequencies (to eliminate foreground sources)
and which have been cross-correlated with other measurements
(Fig.~6).
In the next few years, there will be a sequence of ground-
and balloon-based experiments culminating in
 the NASA MAP (Microwave
 Anisotropy Probe) and the ESA PLANCK satellite missions, which will
 produce all-sky temperature maps with
 a few arcminutes resolution.
These improved maps 
will do much more than measure the position
of the first acoustic peak and, thereby, the curvature; by
measuring the detailed shape of the plateau and a sequence of
peaks with very high precision, they will confirm (or refute) the
basic underlying cosmological scenario and, if confirmed, will
help to determine additional cosmological parameters, such as
$\Omega_m$, $\Omega_{\Lambda}$, $\Omega_b$, $H_0$, 
and more \cite{Kamion,Bond}.
The best-fit parameter region derived from the current CMB results
 shown in Fig. 2,  is consistent with a flat universe, although the
 the  uncertainty is large.  
This analysis 
assumes that the initial fluctuations are adiabatic, as predicted by 
the standard inflationary theory and as assumed in our four 
models (see discussion above).
If they are not this will be apparent from future CMB observations,
and a different means can be used to extract
the curvature from CMB data.
\\
\\
If the universe is flat and
the matter density is less than the
critical density, then there must be some form of nonclustering
dark energy.  In that case, as discussed in the 
introduction, the only way to form the observed  large-scale
structure is if its pressure is negative, because that guarantees
that its density was negligible in the past when structure 
formed.
This conclusion is consistent with the evidence 
suggesting that the universe
is accelerating, which can only occur with a substantial
negative pressure component.

\begin{center}
{\bf The Cosmic Triangle: Present, Past and Future}
\end{center}

The current state of the universe can be surmised from the 
answers to the three questions posed above. The most
precise measurements of the mass (using clusters), the 
acceleration (using supernovae), and the curvature (using 
the CMB) each  confine the universe to a strip in the 
cosmic triangle plot (Fig. 2).  {\it All three 
strips overlap at the $\Lambda${\sc cdm} model with
 approximately $\Omega_m=1/3$,
$\Omega_{\Lambda}=2/3$,
and $\Omega_k=0$~\cite{Wang99,Perlmutter99}.} 
Zero curvature is consistent with inflation.

The verification and refinement of these conclusions will 
take place in the next few years through 
experiments already underway  and will finally settle
some of the questions that have challenged cosmologists for
most of the 20th century.  However, new cosmological
challenges will take their place.
Establishing inflation as the source of the fluctuations that 
seeded galaxy formation requires tests of the shape, 
gaussianity, and gravitational wave component of the 
primordial 
power spectrum \cite{Steinhardt95,Grav,Polarize1,Polarize}.  
As estimates 
of the cold dark matter density become more precise, it 
becomes even more imperative that its composition be 
identified.
 A host of candidates
are suggested by particle physics models~\cite{ParticleDark}.
The leading candidates at present are 
the axion~\cite{ParticleDark1} and the  lightest,
stable, supersymmetry partner particles, such as the photino and
higgsino~\cite{ParticleDark2}.
(Recent measurements of
atmospheric and solar neutrinos show that the
neutrino has a small mass, but the mass is probably  too small to
be important cosmologically~\cite{Atmosneut}.)

However, it is the acceleration of the universe that raises
 the most provocative and profound issues.
 The
acceleration may be caused by a static, uniform vacuum density
(or cosmological constant)
or by a dynamical form of evolving, inhomogeneous dark energy
(quintessence)~\cite{CDS98,Ratra88}.
Distinguishing between the two cosmologically is important because
it informs us of what kind of new fundamental physics is 
required to explain our universe. 
Promising approaches include
measurements of supernovae, CMB anisotropy and gravitational
lensing~\cite{Perlmutter98,Garnavich98,Wang99,Perlmutter99}. 
Special initial conditions are
required for the vacuum energy possibility
  because it remains constant while 
  the matter density
decreases over 100 orders of magnitude as the universe expands.
In order to have a vacuum   energy density only a factor
of two greater than the matter density today, it  would
have to have been exponentially small compared to the matter density
in the early universe.
A major  motivation for proposing quintessence
is that  its interactions can cause
its  energy to naturally adjust itself to be comparable
to the matter density today without special initial 
conditions~\cite{SteinhardtEtAl98}.

Acceleration also affects our projection for the future 
fate of the universe, which can also be represented in a 
cosmic triangle plot (Fig. 7).
As the universe evolves, $\Omega_m$, $\Omega_k$ and $\Omega_{\Lambda}$
change at
different rates, while maintaining a total value
of unity, according to the sum rule.  
Possible trajectories to the future (Fig. 7) 
show that 
$\Omega_m=1$ is an unstable fixed
point and $\Omega_{\Lambda}=1$ is a stable fixed point.  
If $\Omega_m <1$ 
 today and there is any bit
of added dark energy, then we are ultimately careening towards
a flat 
$\Omega_m \rightarrow 0$ ($\Omega_{\Lambda} \rightarrow 1$)
universe in which the matter is
spreading infinitesimally thinly leaving behind only an inert
vacuum energy.  If the vacuum energy (or quintessence)
is unstable this fate may be averted.

As the current millenium 
ends, the past history
and the present state of the universe are making themselves known.
Determining the long-term fate of the universe will 
require an understanding of the fundamental physics 
underlying the dark energy, one of the grand  challenges
for the millenium to come.

\newpage
\begin{table}
\caption{The basic parameters for the four models considered
in this paper: $\Omega_m$, $\Omega_{\Lambda}$, and
$\Omega_k$ are the ratio of the mass, vacuum energy and curvature
to the critical density. $H_0$ is the Hubble parameter in 
km~s$^{-1}$~Mpc$^{-1}$.  The tilt measures how the
amplitude of the inhomogeneity
in initial density perturbations changes with length scale; 
tilt equal to unity means that the amplitude is scale-invariant,
the inflationary prediction.  
The last column is the age of the universe is billions of years.  The first
model, $\Lambda${\sc CDM}, is in best agreement with observations
(see Fig. 2).
 }
\begin{center}
\begin{tabular}{||c||ccc|ccc||}
\multicolumn{6}{c}{Models and  Parameters} 
   \\[0.8ex]
\hline \hline
 Model    &  $\Omega_m$  &  $\Omega_{\Lambda}$ &  $\Omega_k$ & \ $H_0 $  & tilt  &  age \\
   \hline\hline  
Cosmological Const. ($\Lambda${\sc cdm})   & 1/3 & 2/3 & 0 & 65  & 1 & 14.1 
\\ 
Open ({\sc Ocdm})   & 1/3 & 0 & 2/3   & 65  & 1.3 & 12.0 
\\
Standard ({\sc Scdm})   & 1 & 0 & 0  & 50  & 1 & 13.0 \\
Tilted  ({\sc Tcdm})   & 1  & 0 & 0 & 50   & 0.7 & 13.0 \\
 \hline \hline
    \end{tabular}
    \end{center}
    \end{table}

\newpage

  \begin{figure}
 \centerline{\epsfig{file=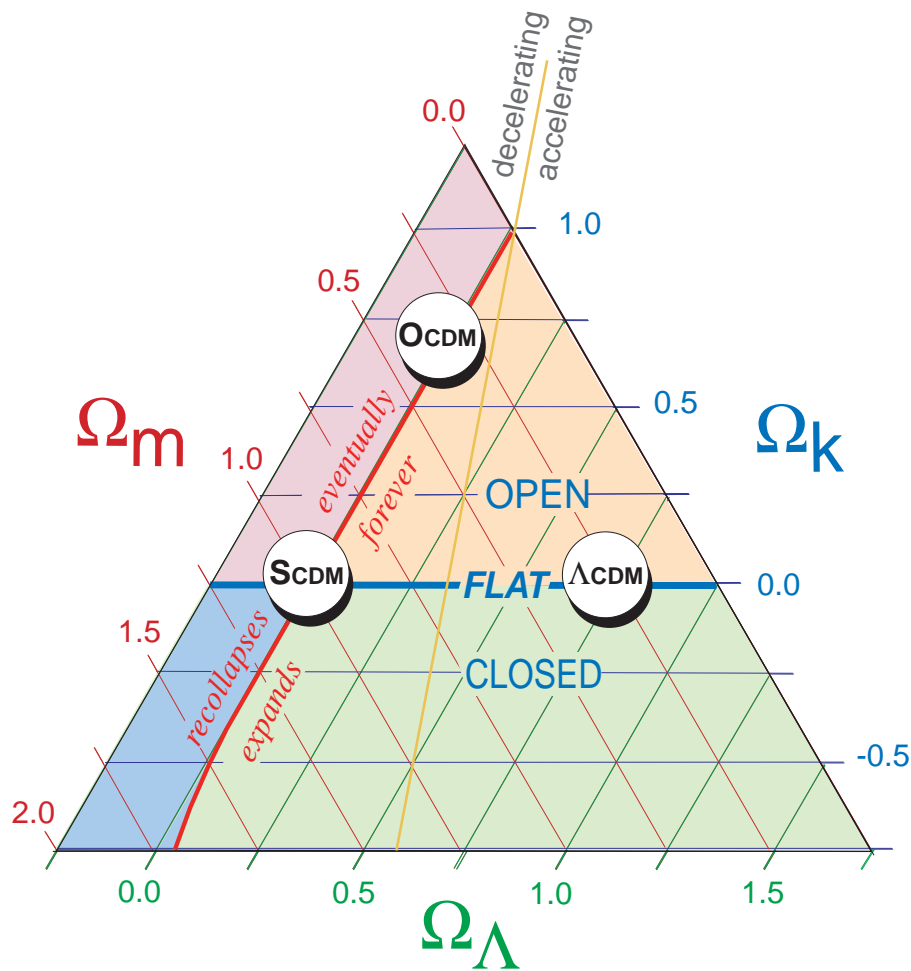,width=5.1in}}
\caption{{\bf The Cosmic Triangle} represents the three
key cosmological parameters 
-- $\Omega_m$, $\Omega_{\Lambda}$,
and $\Omega_k$ --  where
each point in the
triangle 
satisfies the sum rule $\Omega_m + \Omega_{\Lambda} + \Omega_k = 1$.
The blue horizontal line
(marked Flat) corresponds to a flat universe ($\Omega_m +
\Omega_{\Lambda} = 1$), separating
 an open universe from a closed one.
The red line, nearly along the  $\Lambda=0$ line, separates a
universe that will
expand forever (approximately $\Omega_{\Lambda}>0$) from one that will
eventually
recollapse (approximately $\Omega_{\Lambda}<0$). And the yellow, nearly
vertical line
separates a universe with an expansion rate that is currently
decelerating
from one that is accelerating.  The location of three key models are
highlighted: standard cold-dark-matter
({\sc Scdm}) is dominated by matter ($\Omega_m=1$) and no curvature or
cosmological constant;
flat ($\Lambda${\sc cdm}), with $\Omega_m=1/3$, $\Omega_{\Lambda}=2/3$, and
$\Omega_k=0$; and Open CDM ({\sc Ocdm}), with $\Omega_m=1/3$,
$\Omega_{\Lambda}=0$ and curvature
$\Omega_k=2/3$. (The variant, tilted {\sc Tcdm} model is identical in its
position to {\sc Scdm}). 
}
 \end{figure}

 \newpage
  \begin{figure}
\centerline{ \epsfig{file=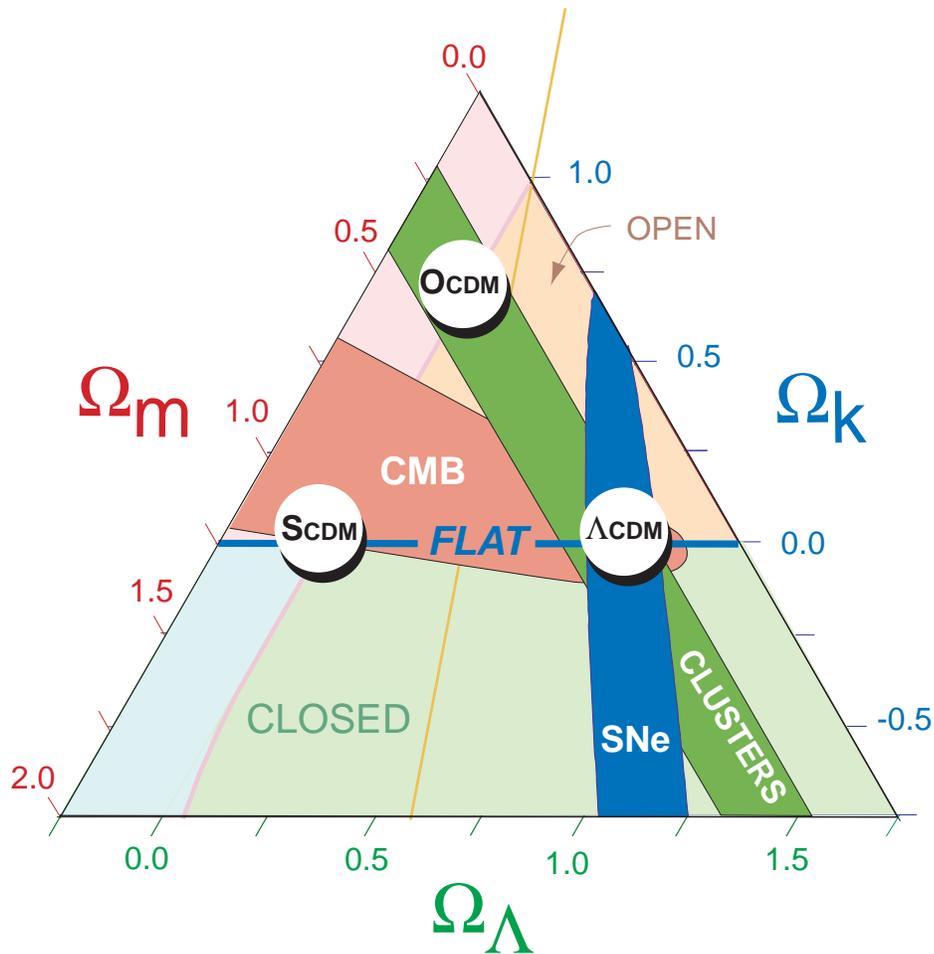,width=5.1in}}
\caption{{\bf The Cosmic Triangle Observed} represents current 
observational 
constraints. The tightest constraints from measurements at
low red shift 
(clusters, including the mass-to-light method, baryon fraction, and cluster
abundance evolution), 
intermediate red shift (supernovae),
and high red shift (CMB) are shown by 
the three color bands (each representing 1-$\sigma$ uncertainties).
Other tests discussed in the paper are consistent with but less
constraining than the constraints illustrated here.
The cluster constraints indicate  a  low-density universe;
the supernovae constraints indicate an accelerating universe; and the
CMB measurements indicate a flat universe.
 The three independent bands intersect 
 at a flat model with
$\Omega_m\sim 1/3$ and $\Omega_{\Lambda}=2/3$; the model
 contains a cosmological constant or other dark energy.} 
 \end{figure}

\newpage
  \begin{figure}
 \centerline{\epsfig{file=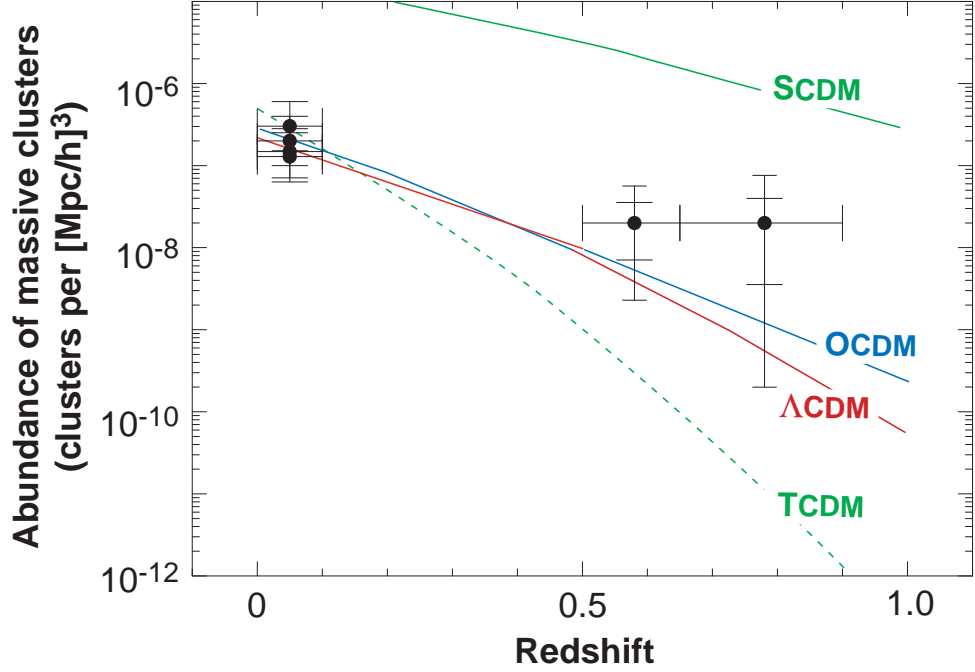,width=5.1in}}
\caption{{\bf The evolution of cluster abundance} 
as a function of red shift is compared with observations~\cite{Bahcall98}
for massive clusters (above $10^{15}$~solar masses within
a 2 Mpc radius, assuming $H_0=65$~km~s$^{-1}$~Mpc$^{-1}$).
Only the   $\Lambda${\sc cdm} and {\sc Ocdm} fit well  the observed
cluster abundance at $z\sim
0$, although the {\sc Tcdm} fits much better than  
the {\sc Scdm} model. See also Fig. 4.
 All four models are normalized to the cosmic microwave background
fluctuations on
large scales (see text).  The observational data points
\protect{\cite{Bahcall98}} (with 1- and
2-$\sigma$ error-bars) show
only a slow evolution in the cluster abundance, consistent with low
$\Omega_m$ models  and inconsistent with  
$\Omega_m=1$.}
 \end{figure}

 \newpage
  \begin{figure}
\centerline{\epsfig{file=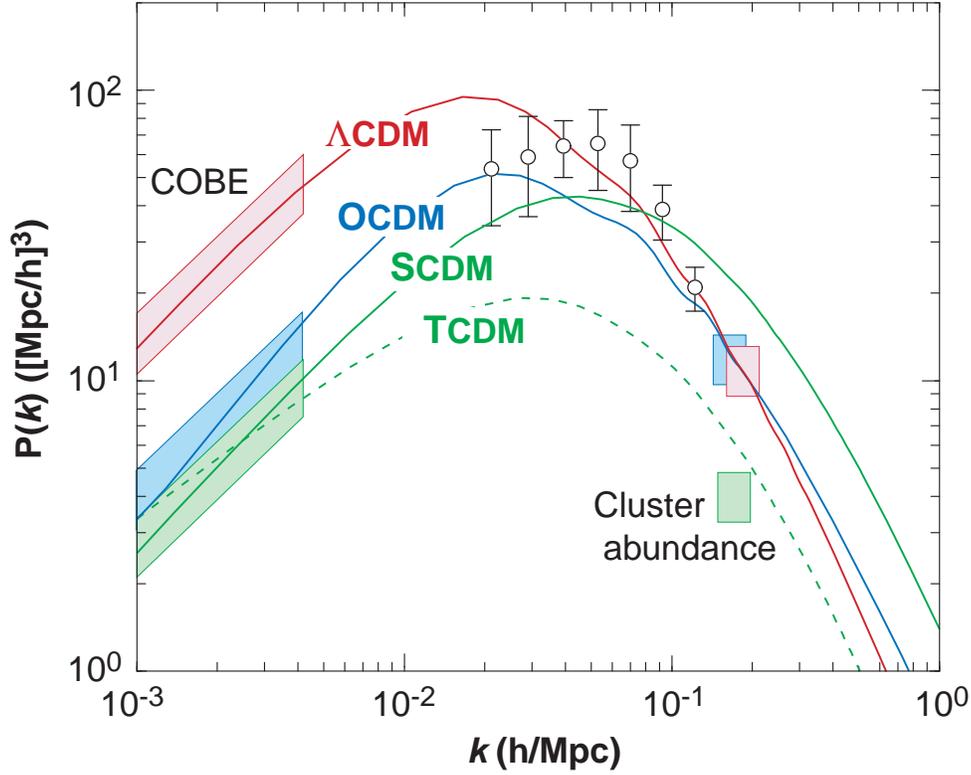,width=5.1in}}
\caption{{\bf The mass power spectrum} represents
the degree of inhomogeneity in the mass distribution
as a function of wavenumber, $k$.
 (The wavenumber is inversely proportional to the length scale;
small scales are to the
right (large k's), and large scales are to the left (small k's)). 
COBE measurements of the CMB anisotropy (boxes on left) and 
measurements of cluster abundance at $z \sim 0$ (boxes
on right) impose different quantitative constraints
for each model;  the  constraints
have been color-coded to indicate the model to which they apply.
All curves are normalized to the CMB fluctuations on large
scales ({\it i.e.,} 
curves are forced to  pass through the COBE error 
boxes on left).
Note that the COBE-normalized 
{\sc Scdm} model significantly overshoots the 
cluster constraint
(green box on right).
The  data points with open circles and 1$\sigma$ error bars
represent the APM galaxy red shift survey~\protect\cite{Peac97}; 
if one assumes bias, then 
this set of  points can be shifted downwards to match the model, 
but the shape of the spectrum suggested by the data is  unchanged.
}
 \end{figure}

 \newpage
  \begin{figure}
\centerline{\epsfig{file=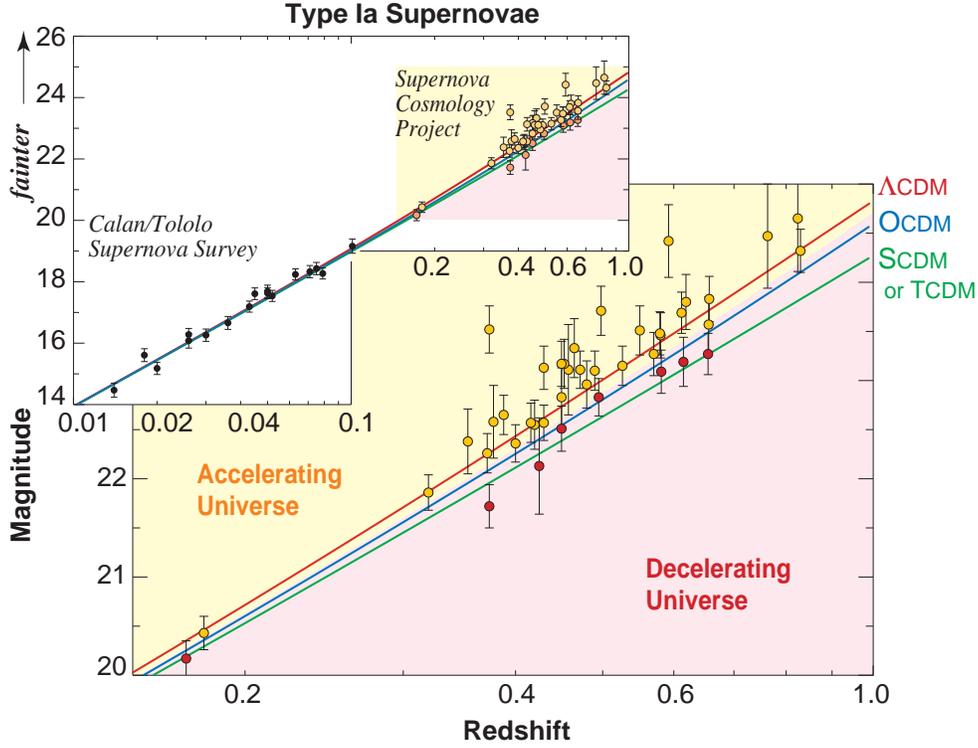,width=5.1in}}
\caption{{\bf Supernovae as standard candles}: 
the relation of observed brightness
(in logarithmic units of ``magnitude'') vs. red shift
for Type Ia supernovae observed at low red shift by
the Calan-Tololo Supernova Survey
and at high red shift by the Supernova Cosmology
Project is presented (with 1$\sigma$ error bars)
 and compared with model expectations.
(Brighter is down and dimmer is up.)  
(All $\Omega_m=1$ flat models yield identical predictions
in this method, thus {\sc Tcdm} is identical to {\sc Scdm}.)
The strong gravitational pull exerted by $\Omega_m=1$ models
(such as {\sc Tcdm} or {\sc Scdm}),
decelerates the expansion rate of the universe and produces 
an apparent `brightening' of high red shift SNIa, whereas  
the effect of a cosmological constant accelerating the expansion 
rate (as in $\Lambda${\sc cdm}) is seen as a relative `dimming' of
the distant SNIa caused by their larger distances. 
 The lower-right plot shows a
close-up view of the expected deviations between the three
models as a function of red shift.  The background color
(and shading of the data points) indicates the region for
which the universe's expansion would accelerate (yellow)
or decelerate (red)  
for $\Omega_m \sim 0.2$.  (Higher
values of $\Omega_m$ would extend the yellow
accelerating-universe region farther down on this plot.)
Similar results are found by the HZS team \protect{\cite{Riess98}},
as discussed in the text.
The results
provide  evidence  for an accelerating expansion rate.
}
 \end{figure}

 \newpage
  \begin{figure}
\centerline{\epsfig{file=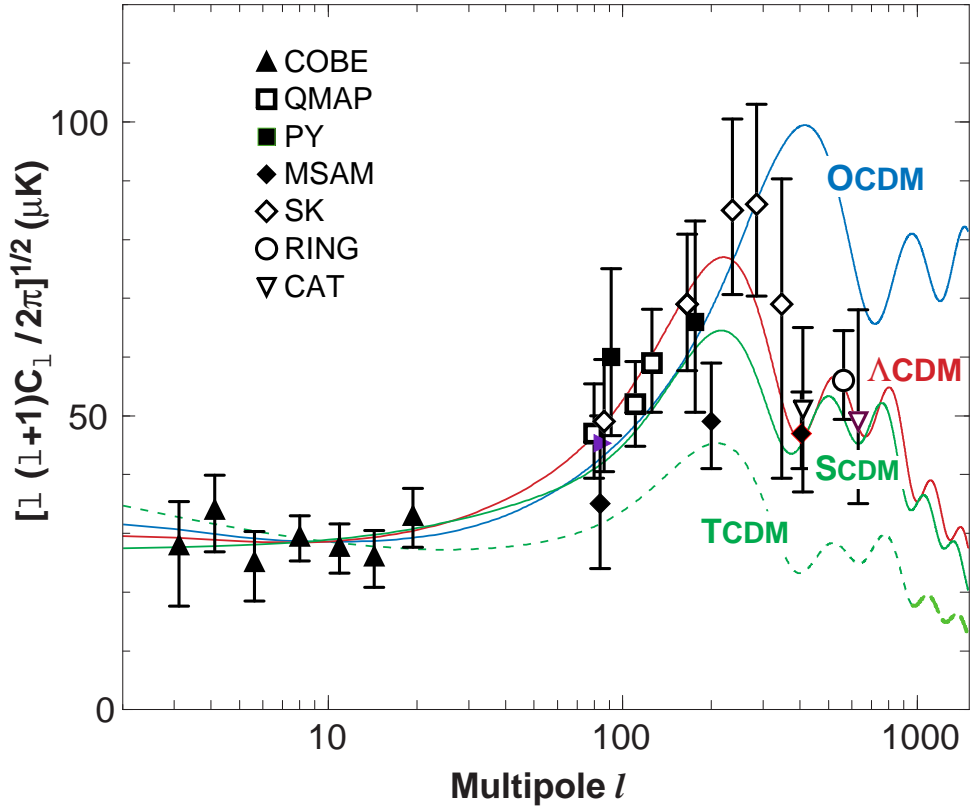,width=5.1in}}
\caption{{\bf The cosmic microwave background temperature anisotropy}
is presented as a function of angular scale.  The multipole
$\ell$ corresponds roughly to an angular scale of $\pi/\ell$ radians.
Flat models ($\Omega_m + \Omega_{\Lambda}=1$)
produce an acoustic peak at $\ell=\sim 200$ (about one degree on the sky). Open
models have a peak that is shifted to smaller scales
(larger $\ell$'s). (The height of the peak depends on additional
parameters, including $\Omega_m$, $\Omega_{\Lambda}$, $\Omega_b$, $H_o$,
tilt;
here we use the model  values from Table I.)  The observational data points
(with 1$\sigma$ error bars)
include the
COBE measurements on large scales (small $\ell$'s) and other published,
multi-frequency ground- and balloon-based observations (see text for
references).}
 \end{figure}

 \newpage
 \begin{figure}
 \centerline{\epsfig{file=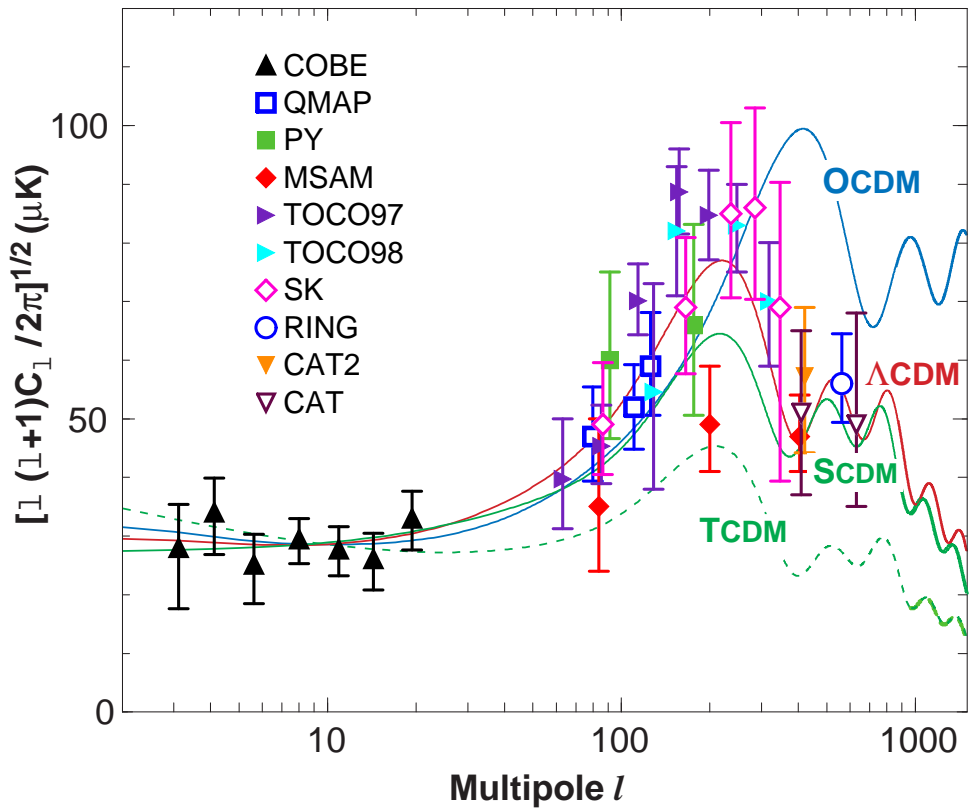,width=5.1in}}
 \caption{ Same as previous figure except with data from 
 recent preprints added.
 }
  \end{figure}

 \newpage
  \begin{figure}
\centerline{ \epsfig{file=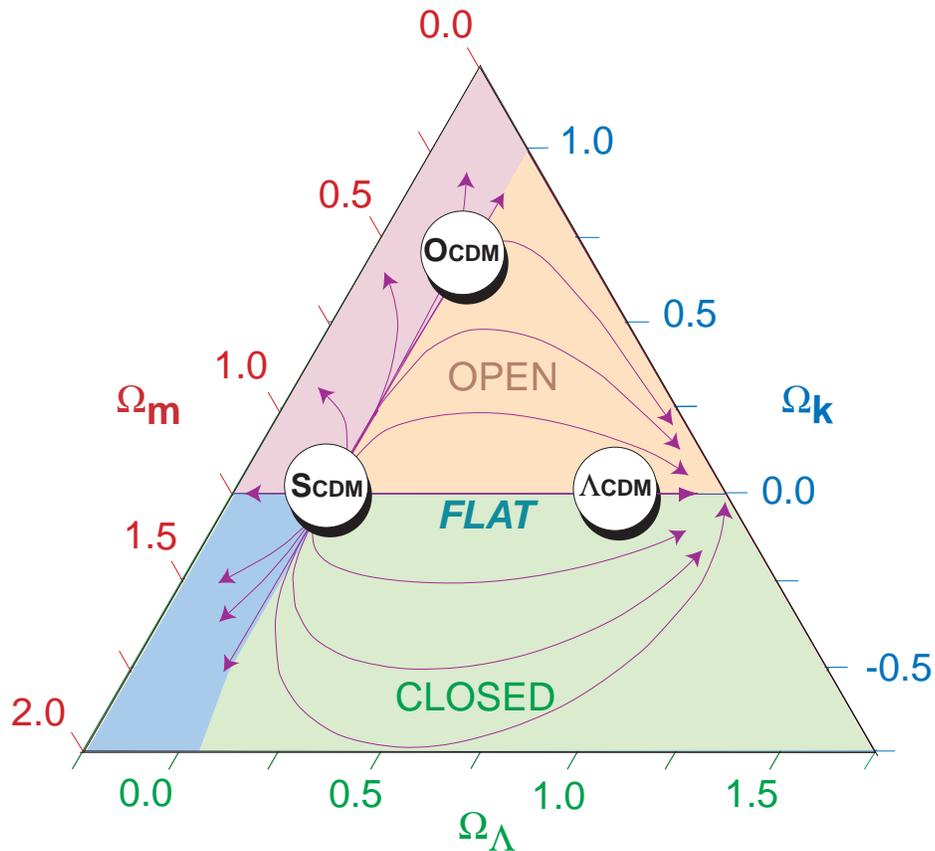,width=5.1in}}
\caption{{\bf The past and  future of the universe} are represented by various
trajectories in the cosmic triangle plot.  The trajectories, which
originate
from near $\Omega_m=1$ (an unstable equilibrium point matching the
approximate condition of the universe during early structure formation),
indicate the  path traversed in the triangle plot as 
the universe evolves.  For the current best-fit
$\Lambda${\sc cdm}
model, the future represents a flat, accelerating universe that expands
forever, ultimately reaching $\Omega_m \rightarrow 0$ and
$\Omega_\Lambda
\rightarrow 1$. }
 \end{figure}


\begin{thebibliography}{}
\bibitem{footcold} "Cold" matter means
that the kinetic energy is much less than the mass energy.
Non-baryonic
 means that the matter is not made of 
of ``ordinary'' matter
such as protons and neutrons (called baryons by particle
physicists).
\bibitem{footnote1} 
A simple computation
 in general relativity shows that the ratio of the dark energy density to the
 matter density is $a(t)^{-3p/\rho}$ where $p$ is the pressure, $\rho$
 is the energy density of the dark energy, and $a(t)$ is the
 scale factor of the universe at a given time.  As one extrapolates back
 in time in an expanding universe, the scale factor $a$ gets smaller.
 So, if $p > 0$ the ratio increases in the past, the matter density
 would never have dominated, and structure would never have formed.
 If $p < 0$ (negative pressure) the ratio decreases in the past;
 the matter density would have dominated, and structure would have formed.
\bibitem{over}
The vacuum energy can overtake the matter density and curvature and 
prevent recollapse in a closed universe
because the different components
evolve differently with time: the matter density decreases
as $1/a^3$; the curvature decreases as $1/a^2$; and the cosmological
constant remains constant, thus, eventually dominating the 
energy of the
universe.)
\bibitem{foot30}
It is conventional to write $\Omega_i= \rho_i/\rho_c$, where $\rho_c$
is the critical density, $\rho_c
\equiv 3 H^2/8 \pi G = 1.7 \times 10^{-29}$~g/cm$^3$ (for the present 
value of $H_0=65$~km~s$^{-1}$~Mpc$^{-1}$). Note that $\rho_c$ is 
proportional to $H^2$, which decreases with time.
\end{thebibliography}
\end{document}